\documentclass{nature}

\usepackage[utf8]{inputenc}
\usepackage[T1]{fontenc}

\usepackage{color}
\usepackage{caption}

\usepackage{graphicx}
\usepackage{verbatim}
\usepackage{xcolor}
\usepackage{amsmath}
\usepackage{upgreek}
\newcommand{\beginsupplement}{%
        \setcounter{table}{0}
        \renewcommand{\thetable}{S\arabic{table}}%
        \setcounter{figure}{0}
        \renewcommand{\thefigure}{S\arabic{figure}}%
     }

\makeatletter
\newenvironment{figurehere}
{\def\@captype{figure}}
{}
\makeatletter

\makeatletter
\let\saved@includegraphics\includegraphics
\AtBeginDocument{\let\includegraphics\saved@includegraphics}
\renewenvironment*{figure}{\@float{figure}}{\end@float}
\makeatother

\title{Memristive control of mutual SHNO synchronization for neuromorphic computing}

\author{Mohammad Zahedinejad$^{1,2}$, Himanshu Fulara$^{1}$, Roman Khymyn$^{1}$, Afshin Houshang $^{1,2}$, \\Shunsuke Fukami$^{4,5,6,7,8}$, Shun Kanai$^{4,5,6,9}$, Hideo Ohno$^{4,5,6,7,8}$, \& Johan \AA kerman$^{1,2,3}$}

\begin{document}

\maketitle

\begin{affiliations}
 \item Physics Department, University of Gothenburg, 412 96 Gothenburg, Sweden
 \item NanOsc AB, Electrum 229, 164 40 Kista, Sweden
 \item Material and Nanophysics, School of Engineering Sciences, KTH Royal Institute of Technology, Electrum 229, 164 40 Kista, Sweden
 \item Laboratory for Nanoelectronics and Spintronics, Research Institute of Electrical Communication, Tohoku University, 2-1-1 Katahira, Aoba-ku, Sendai 980-8577, Japan
  \item Center for Science and Innovation in Spintronics, Tohoku University, 2-1-1 Katahira, Aoba-ku, Sendai 980-8577 Japan
  \item Center for Spintronics Research Network, Tohoku University, 2-1-1 Katahira, Aoba-ku, Sendai 980-8577 Japan
 \item Center for Innovative Integrated Electronic Systems, Tohoku University, 468-1 Aramaki Aza Aoba, Aoba-ku, Sendai 980-0845 Japan
 \item WPI-Advanced Institute for Materials Research, Tohoku University, 2-1-1 Katahira, Aoba-ku, Sendai 980-8577 Japan
  \item Division for the Establishment of Frontier Sciences, Tohoku University, 2-1-1, Katahira, Aoba-ku, Sendai 980-8577 Japan

\end{affiliations}

\begin{abstract}

Synchronization of large spin Hall nano-oscillators (SHNO) arrays is an appealing approach toward ultra-fast non-conventional computing based on nanoscale coupled oscillator networks. However, for large arrays,  interfacing to the network, tuning its individual oscillators, their coupling, and providing built--in memory units for training purposes, remain substantial challenges. Here, we address all these challenges using memristive gating of W/CoFeB/MgO/AlO$_x$ based SHNOs. In its high resistance state (HRS), the memristor modulates the perpendicular magnetic anisotropy (PMA) at the CoFeB/MgO interface purely by the applied electric field. In its low resistance state (LRS), and depending on the voltage polarity, the memristor adds/subtracts current 
to/from the SHNO drive. 
The operation in both the HRS and LRS affects the SHNO auto--oscillation mode and frequency, which can be tuned 
up to 28~MHz/V. 
This tuning allows us to reversibly turn on/off mutual synchronization in chains of four SHNOs. We also demonstrate two individually controlled memristors 
to tailor both the coupling strength and the frequency of the synchronized state.  Memristor gating is therefore an efficient approach to input, tune, and store the state of the SHNO array for any non-conventional computing paradigm, all in one platform.

\end{abstract}

\section*{Introduction}

The demand for compact communication systems and faster processing hardware is booming as we approach the era of the Internet of Things (IoT). However, with the approaching end of Moore's law and the increasing demand for mobile applications, it seems unlikely that 
state-of-the-art CMOS technologies can live up to all IoT expectations.\cite{shalf2015computing} Emerging paradigms---such as machine vision and cognitive tasks for connected objects---have severely challenged the existing technologies in terms of power consumption and processing speed, tasks that the human brain, thanks to its massive neuronal connectivity, performs extremely fast using negligible power compared to conventional processors.\cite{buzsaki2006rhythms} Although there have been efforts to imitate the neuronal connectivity by involving many transistors representing individual neurons, Von--Neumann computing inherently does not include connected and interactive computing elements.\cite{hsu2014ieee} Numerous types of unconventional computing hardware have been proposed by different research communities to address such von--Neumann shortcomings, each outperforming CMOS processors in some aspects: Memristive computing\cite{boybat2018ntcom,wang2018nt}, optical computing\cite{inagaki2016science,solli2015nt,o2007sci}, spintronic computing\cite{borders2016ape,torrejon2017nt,grollier2020nt,manipatruni2018nt}, and quantum computing\cite{clarke2008nt,ladd2010nt} are only a few examples of such hardware.  

Using coupled networks of oscillators as an alternative computing paradigm has been proposed since many different technologies can explore this route, ranging from mechanical oscillators\cite{fang2016SciAdv,shim2007sci} through memristive\cite{kumar2017nt,ignatov2017sci} and superconducting oscillators\cite{segall2017pre,galin2018pra,inagaki2016science} to optical\cite{mcmahon2016science}, and spintronic\cite{torrejon2017nt,romera2018nt,yogendra2016ieee,awad2017natphys} oscillators. Among these, the spintronic oscillator is a promising candidate meeting all technical requirements, such as room--temperature operation, true miniaturization, CMOS integration, high-speed, and low power consumption. In a recent study, a short chain of four weakly coupled spin transfer torque nano--oscillators (STNOs) was used to perform vowel recognition.\cite{romera2018nt} Through tuning of the individual STNO currents, and their injection phase--locking to two external microwave sources, a recognition rate comparable to state-of-the-art CMOS processors was achieved. However, more complex tasks will require many more oscillators, and the demonstrated STNOs do not offer an attractive scaling path to much larger networks. We have recently demonstrated two--dimensional (2D) synchronization in large arrays of an alternative emerging class of nanoscale spin-based oscillators---so-called spin Hall nano--oscillators (SHNOs).\cite{zahedinejad2020nt} Arrays accommodating up to 64 strongly coupled SHNOs, operating at 10 GHz, achieved robust synchronization and a corresponding record-high quality factor of 170,000. We also demonstrated that 
the 2D--SHNO arrays were capable of performing the same type of computation as STNO chains\cite{romera2018nt} while they operate at a 25 times higher frequency. Adding the potential to control individual SHNOs using direct voltage control\cite{fulara2020nt}, such SHNO arrays offer a viable scaling path to very large nano-oscillator networks of SHNOs that can be scaled down to 20 nm\cite{durrenfeld2017nanoscale}. 

In addition, most oscillator network computation proposals require not only interacting oscillators but also 
on-chip memory for training purposes.
\cite{parihar2017scirep,velichko2019elec,zhang2019elec,kumar2017scirep,nikonov2015ieee,fang2016SciAdv,fang2017acm,novikov2015proced,wang2019proced} There is hence both a practical and computational need for non-volatile storage of individual oscillator properties, a challenge that remains unsolved for nanoscale arrays.

In this work we present an ultra-low-power path to controlling the magnetization dynamics in W/(Co$_{0.75}$Fe$_{0.25})_{75}$B$_{25}$/MgO based SHNOs by applying an electric field via gating. The use of electric field (voltage) to control magnetism has been extensively studied in the literature, \emph{e.g.}~using 
voltage controlled magnetic anisotropy (VCMA) to achieve low power switching in magnetic tunnel junction (MTJ) memory cells by controlling/modulating the perpendicular magnetic anisotropy (PMA)\cite{shiota2012nt,wang2012natmat,kanai2012apl,kanai2014apl,kanai2016apl,zhang2016scirep}. As PMA can have a significant impact on the operating frequency and the auto-oscillation (AO) mode profile of W/CoFeB/MgO based SHNOs,\cite{zahedinejad2018apl,fulara2019sciadv} our voltage control of the CoFeB/MgO interface 
tunes the local magnetodynamical properties, 
resulting in a change in the precessional angle and the mode volume\cite{fulara2020nt, dvornik2018pra} that define the SHNO frequency and the coupling between SHNOs. In addition to the direct voltage control, we also show that the gates can exhibit a pronounced memristive behavior, which we then use to control both the individual oscillators and their mutual coupling to make them either operate in synchrony or break their synchronization. Synchronization control can be achieved in both the high resistance state (HRS) and the low resistance state (LRS) of the memristor, albeit through different mechanisms: electric field-driven in the HRS, and current-driven in the LRS. The memristors also add embedded memory functionality that can be used to recall the past input values each oscillator was exposed to. The demonstrated 
locally embedded memristive tuning capability of individual SHNOs and their coupling in large arrays provides tremendous added functionality and a direct scaling path towards the realization of 
large nano-scale coupled oscillatory networks for a range of different non-conventional computing paradigms.

\begin{figure}
  \begin{center}
  \includegraphics[width=5.5in]{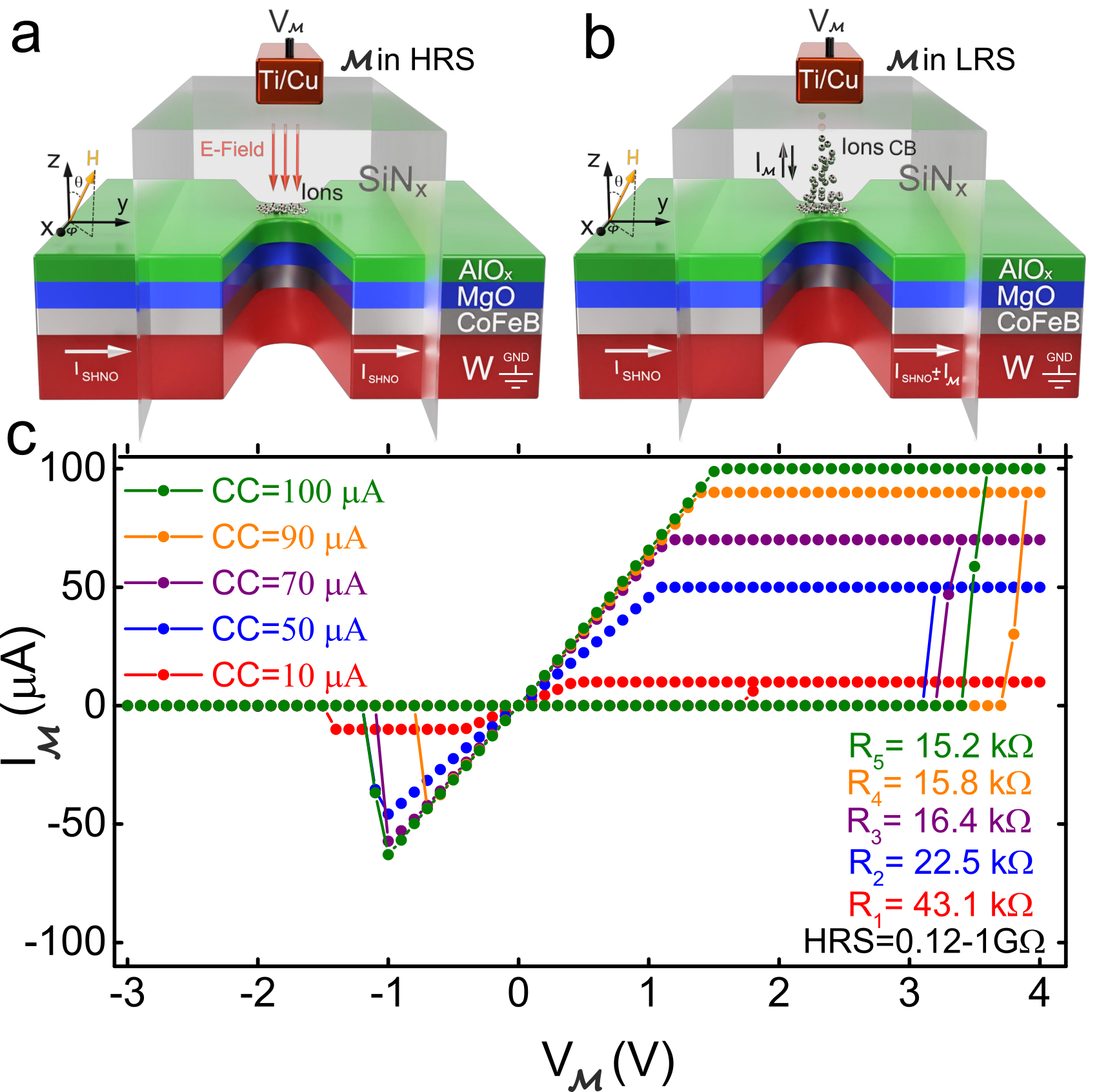}
  \renewcommand{\baselinestretch}{1}
    \caption{\textbf{Device schematic and multi-level resistive switching of the memristor gate.} Schematic of a SHNO with a memristive gate fabricated on top of the nano-constriction sharing a common ground contact when operating in \textbf{(a)} HRS and \textbf{(b)} LRS. A W(5)/(Co$_{0.75}$Fe$_{0.25})_{75}$B$_{25}$(1.7)/MgO (2)/AlO$_x$(2) multilayer (numbers in parentheses indicate thickness in nm) is buried in SiN$_x$(15) with a Ti(2)/Cu(40) top electrode as gate contact. The coordinate system shows the magnetic field vector $\mathrm{\mu_0}H$ applied at an out--of--plane angle  $\theta$ and an in--plane angle $\phi$. Upon sweeping $V_\mathcal{M}$, an electric field driven reversible conductive bridge is formed and retracted through the insulating multilayer and the memristor current $I_\mathcal{M}$ is added/subtracted from $I_\mathrm{SHNO}$ depending on the $V_\mathcal{M}$ polarity. 
    \textbf{(c)} multi-level resistive switching of the memristive gate, controlled by the current compliance of the power supply.}
    \label{fig:1}
    \end{center}
\end{figure}

\section*{Results and discussion}
\subsection{Gated SHNO and memristor geometries.}
\sloppy After defining nano-constriction based SHNOs, to apply an electric field to the SHNO, we covered the entire W(5~nm)/(Co$_{0.75}$Fe$_{0.25})_{75}$B$_{25}$~(1.7~nm)/MgO (2~nm)/AlO$_x$(2~nm) based SHNO (see e.g. Ref.~\cite{zahedinejad2018apl,fulara2019sciadv,fulara2020nt}) with 15~nm SiN$_x$ and placed a 100~nm wide Ti/Cu gate on top of the 120~nm wide nano--constriction region. Fig.~\ref{fig:1}a and b show the schematic of the fabricated SHNO and the memristor in two operating regimes, HRS and LRS. The memristor shares its bottom electrode with the ground contact of the SHNO. The multilevel resistance states obtained by reversible memristive switching pinched at the origin is shown in Fig.~\ref{fig:1}c. The different low resistance states (LRS) of the memristor were achieved by sweeping the gate voltage while keeping the current compliance (CC) of the voltage source to different values; this controls the size of the conducting path formed inside the insulating multilayer (MgO(2~nm)/AlO$_x$(2~nm)/SiN$_x$(15~nm)). In practice, setting the memristor into various resistance states is done by programming the width of voltage pulses applied to the memristor. The (negligible) memristor current $I_\mathcal{M}$ in the high resistance state (HRS) is defined by the leakage current of the insulating multilayer. $I_\mathcal{M}$ in both the HRS and the LRS is added/subtracted to/from the drive current applied to the SHNO ($I_{\mathrm{SHNO}}$) depending on the memristor voltage polarity. Because of the presence of the metal oxide layers and active upper electrode metals, 
MgO,  AlO$_x$ and Ti/Cu, the memristive switching mechanism is believed to be governed by either oxygen or metal ion migration forming a filamentary conduction bridge (CB) inside the insulating multilayer~\cite{zhang2013apl,mohammad2016nanotech} as shown in Fig.\ref{fig:1}a and b. 

\begin{figure}
  \begin{center}
  \includegraphics[width= 5.5in]{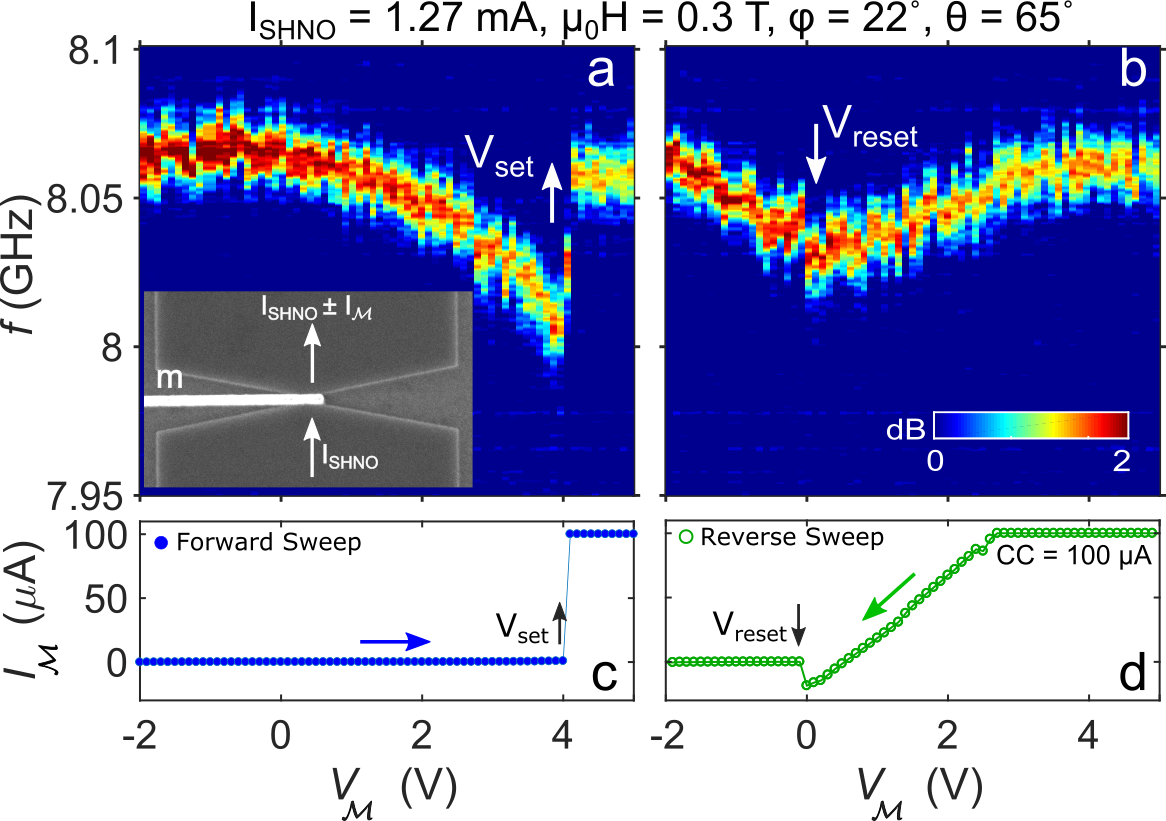}
  \renewcommand{\baselinestretch}{1}
    \caption{\textbf{Memristive control of SHNO auto-oscillations (AO).} $I_{\mathrm{SHNO}}$~= 1.27~mA in all plots. $\mu_0 H$~=~0.3 T was applied at $\theta$~=~65$^{\circ}$ and $\phi$~=~22$^{\circ}$. \textbf{(a)} SHNO frequency tuned by forward sweep of the memristor voltage from its HRS, modifying the PMA by electric field until $V_\mathcal{M}$~=~$V_{\mathrm{set}}$ where the AO is instead controlled by $I_\mathcal{M}$ in the LRS. Inset: SEM image of the gate on top of the SHNO. \textbf{(b)} AO for the reverse $V_\mathcal{M}$ sweep. The memristor returns to its HRS at $V_\mathcal{M}$~=~$V_{\mathrm{reset}}$ and the electric field regains control of the AO. \textbf{(c)} and \textbf{(d)} show the memristor current, $I_\mathcal{M}$, for both forward and reverse sweep.}
    \label{fig:2}
    \end{center}
\end{figure}

Fig.\ref{fig:2} shows the memristive control of the  power spectral density (PSD) for a 120~nm SHNO operating at $I_{\mathrm{SHNO}}$~= 1.27~mA while a magnetic field of $\mu_0 H$~=~0.3 T was applied at an out--of--plane (OOP) angle of $\theta$~=~65$^{\circ}$ and an in--plane (IP) angle of  $\phi$~=~22$^{\circ}$ (see methods and supplementary figure 4 for  detailed description of electrical measurements). The memristor is in the HRS when $V_\mathcal{M}$ is swept in the forward direction from -2~V to 4~V as Fig.\ref{fig:2}c shows a negligible value for $I_\mathcal{M}$. In the HRS the conductive path is not yet formed in the memristor, so the applied voltage induces a strong electric field across the insulating multilayer changing the PMA field ($H_{\text{k}}^{\perp}$) and consequently the effective magnetization described by $M_{\text{eff}}= M_{\text{s}}-H_{\text{k}}^{\perp}$. The change in $M_{\text{eff}}$ directly translates into a change in the precession angle of the magnetization, resulting in an overall 

\noindent 60 MHz (10~MHz/V) AO frequency change as shown in Fig.\ref{fig:2}a. When the memristor switches to one of its LRSs, in this case defined by CC $=$ 100~$\mu$A, at $V_\mathcal{M}=V_{\mathrm{set}}=$ 4 V, a non-negligible $I_\mathcal{M}$ is added to $I_{\mathrm{SHNO}}$ which in turn sharply increases the magnitude of spin current. The increase of spin current reflects in operating frequency of the SHNO as it is experimentally observed by a jump in the SHNO frequency. The frequency jump is associated with the memristor resistance state and each LRS defines a new frequency state. In the reverse sweep shown in Fig.\ref{fig:2}b and d, the AO frequency linearly follows $I_\mathcal{M}$ until $V_\mathcal{M}$ reaches $V_{\mathrm{reset}}$ and $I_\mathcal{M}$ again becomes negligible. When the SHNO chain is biased, the memristor $I - V$ plot is no longer pinched at 0~V since the bottom electrode is at a non-zero potential. From $V_{\mathrm{reset}}$, the memristor is back to its HRS where the AO frequency is controlled by the electric field. One fundamental importance of the memristive gating hence lies in the fact that the memristive mechanism provides two qualitatively very different regimes with their respective tuning knobs for the frequency of the SHNO---an electric field and a memristor current. 

To study the potential for using both the electric field and the memristor current to tune individual nano-constrictions in SHNO arrays, we fabricated a chain accommodating two SHNOs and placed two memristive gates, $\mathcal{M}_1$ and $\mathcal{M}_2$, on top of the nano-constriction regions as shown in the inset in Fig.\ref{fig:3}a. When both gates are floating, \emph{i.e.}~no voltage applied to either $\mathcal{M}_1$ and $\mathcal{M}_2$, sweeping $I_{\mathrm{SHNO}}$ brings the two SHNOs into synchronization at about $I_{\mathrm{SHNO}}$~=~0.8~mA as shown in Fig.\ref{fig:3}a. The SHNOs stay synchronized until $I_{\mathrm{SHNO}}$ reaches 0.9~mA. As our initial state, we therefore fix $I_{\mathrm{SHNO}}$ to 0.85~mA corresponding to a stable synchronized state. 

Fig.\ref{fig:3}b shows the memristive switching of $\mathcal{M}_1$ as a function of the applied gate voltage, $V_{\mathcal{M}_1}$. Starting from $V_{\mathcal{M}_1}=$ --2 V, $\mathcal{M}_1$ stays in its HRS until it switches to its LRS at $V_{set}=$ 2.4 V (blue solid data points and arrow). Upon reversing the voltage sweep direction, $\mathcal{M}_1$ stays in its LRS until it is reset at about --0.2 V (green hollow data points and arrow). The corresponding response

 \begin{figurehere}
  \begin{center}
  \includegraphics[width=5.5in]{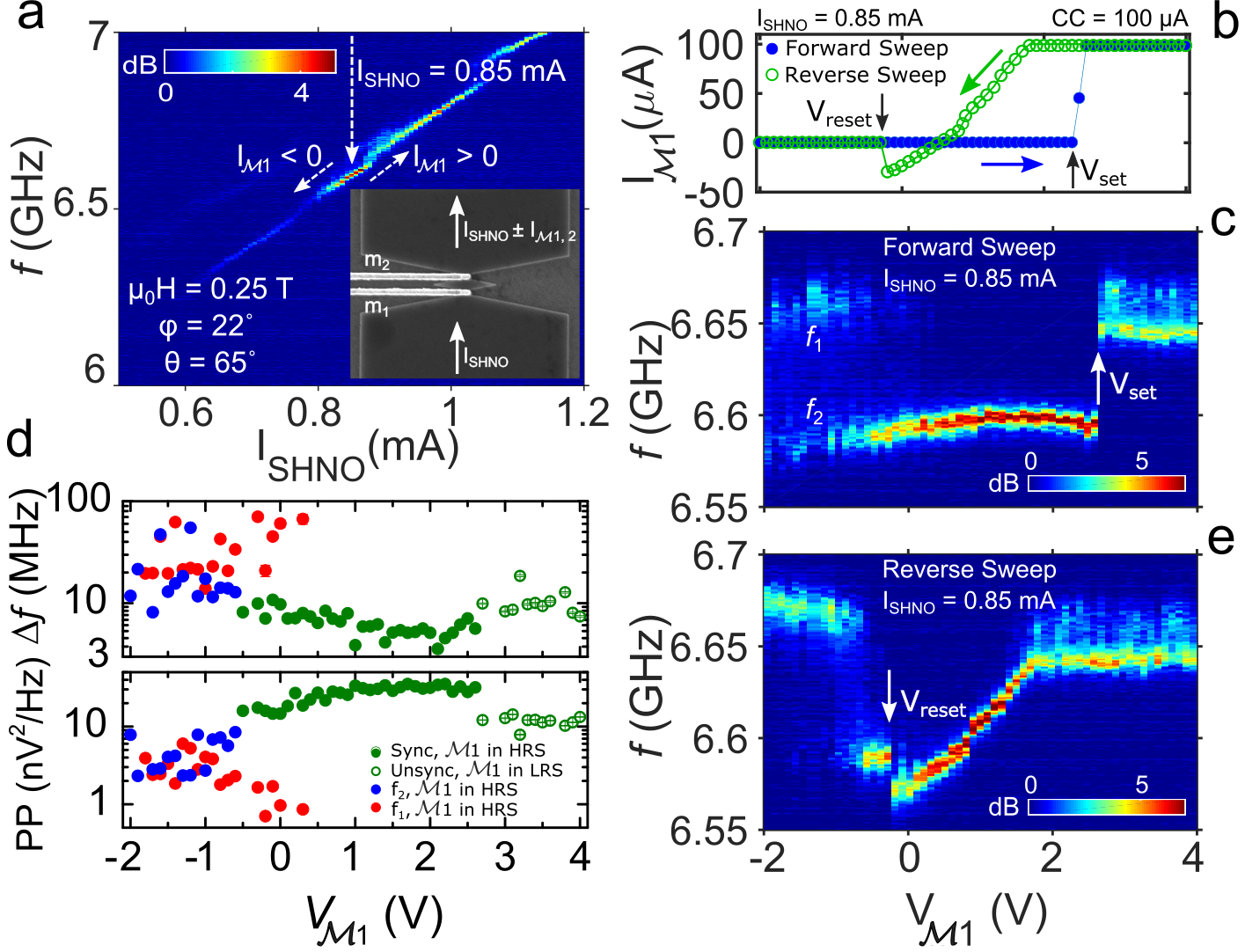}
  \renewcommand{\baselinestretch}{1}
    \caption{\textbf{A chain of two SHNOs with two memristive gates. 
    } \textbf{(a)} PSD \emph{vs.}~$I_{\mathrm{SHNO}}$ in a field of $\mu_0 H$~=~0.25 T applied at $\theta$~=~65$^{\circ}$ and $\phi$~=~22$^{\circ}$. The AO shows a synchronized region for a small $I_{\mathrm{SHNO}}$ range around 0.85~mA. The inset shows an SEM image of the two nano-constrictions with their respective memristor gates, $\mathcal{M}_1$ and $\mathcal{M}_2$. 
    \textbf{(b)} $I_{\mathcal{M}_1}$ \emph{vs.} $V_{\mathcal{M}_1}$ showing reversible switching between HRS and LRS defined by CC~=~100 $\mu$A for a constant $I_{\mathrm{SHNO}}$~=~0.85~mA. \textbf{(c)} PSD \emph{vs.} $V_{\mathcal{M}_1}$ for the forward sweep. The electric field induced by $V_{\mathcal{M}_1}$ in the HRS breaks the synchronized state and recovers it again as $\mathcal{M}_1$ approaches $V_{\mathcal{M}_1}$~=~$V_{\mathrm{set}}$. In the LRS, an $I_{\mathcal{M}_1}$~=~+100 $\mu$A adds to the $I_{\mathrm{SHNO}}$ increasing the AO to a non-synchronized frequency in \textbf{(a)}.
    \textbf{(d)} Extracted peak power and linewidth values \emph{vs.} $V_{\mathcal{M}_1}$ for the forward sweep in \textbf{(c)}. \textbf{(e)} PSD \emph{vs.}~$V_{\mathcal{M}_1}$ for reverse sweep. The AO frequency linearly follows $I_{\mathcal{M}_1}$ recovering the synchronized state until $\mathcal{M}_1$ switches back to its HRS and synchronization breaks apart again.}
    \label{fig:3}
    \end{center}
\end{figurehere}

\noindent of the SHNO chain is shown in Fig.\ref{fig:3}c. We first note that applying a negative voltage to $\mathcal{M}_1$ breaks the synchronized state into two distinct frequency branches,$f_{1}$ and $f_{2}$, with lower peak power (PP) and larger linewidth ($\Delta f$) values as shown by the red and blue solid data points in Fig.\ref{fig:3}d. 
As $V_{\mathcal{M}_1}$ approaches zero, 
the synchronized state is retrieved and appears to grow more robust since both PP and $\Delta f$ continues to improve. However, once $\mathcal{M}_1$ switches to its LRS, a substantial $I_{\mathcal{M}_1}$~=~100 $\mu$A ($I_{\mathcal{M}_1}$ > 0) is sharply injected into the entire chain causing a jump in the operating frequency, as shown by the white arrow in Fig.\ref{fig:3}a. At this operating point, the chain is no longer robustly synchronized 
and both PP and $\Delta f$ deteriorate, as shown by hollow green data points in Fig.\ref{fig:3}d. However, as Fig.\ref{fig:3}e depicts, when $I_{\mathcal{M}_1}$ is reduced during the reverse sweep, the synchronized state is once again recovered even when $I_{\mathcal{M}_1}$ eventually changes its sign ($I_{\mathcal{M}_1}$ $<$ 0) and opposes $I_{\mathrm{SHNO}}$ while $V_{\mathcal{M}_1}$ crosses the pinch off voltage. Finally, the synchronization again breaks  when $\mathcal{M}_1$ is back to the HRS at $V_{\mathcal{M}_1}$~=~$V_{\mathrm{reset}}$. A very similar behavior was observed when $V_{\mathcal{M}_2}$ forward and reversed sweeps were performed on $\mathcal{M}_2$ (see Supplementary Figure 1).

We then studied the electric field and memristive tuning on the coupling within a chain accommodating four SHNOs in series. Two memristive gates, $\mathcal{M}_1$ and $\mathcal{M}_2$, were again fabricated but this time placed on top of the bridges connecting the SHNOs 2~\&~3 and 3~\&~4 as shown in the inset in Fig.\ref{fig:4}a. As the connecting bridges are the regions where the AO modes can overlap and interact, we intend to modify the synchronization properties also this way. The PSD profile of the chain vs.~$I_{\mathrm{SHNO}}$ (both gates floating), shown in Fig.\ref{fig:4}a, indicates that the four oscillators start in a synchronized state 
and later break apart into two frequency branches for larger $I_{\mathrm{SHNO}}$. A value of $I_{\mathrm{SHNO}}$~=~0.712~mA (dotted white arrow in Fig.\ref{fig:4}a), at which the synchronization is broken
, was chosen as the operating point to conduct the subsequent gating operation. In the HRS (indicated by the small $I_{\mathcal{M}_1}$ Fig.\ref{fig:4}b), sweeping $V_{\mathcal{M}_1}$ forward (while $\mathcal{M}_2$ is floating) brings the upper frequency branch down towards the lower frequency branch as shown in Fig.\ref{fig:4}c. At $V_{\mathcal{M}_1}$~=~$V_{\mathrm{sync}}$ (white arrow in Fig.\ref{fig:4}c) and until $V_{\mathcal{M}_1}$~=~4.2~V, the two frequency branches continue to operate in unison as also indicated by the 50\% drop of the linewidth  and the significant increase in the peak power (see Supplementary Figure 2). 
As the VCMA induced by $V_{\mathcal{M}_1}$ increases the weak coupling to recover the broken synchronization, the chain remains synchronized even when $\mathcal{M}_1$ switches to the LRS and $I_{\mathcal{M}_1}$~=~100~$\mu$A adds 
to SHNO 3~\&~4. In the HRS, when 
$V_{\mathcal{M}_1}$ is swept from --3 to +4.2~V, an overall frequency change of 203~MHz is achieved, corresponding to a tunability of 28~MHz/V. 
For the reverse sweep direction presented in Fig.\ref{fig:4}d, the synchronization state is extended for a larger range of $V_{\mathcal{M}_1}$  down to $V_{\mathrm{un-sync}}$ (white arrow in Fig.\ref{fig:4}d) as the oscillators tend to pull each other's frequencies when they are synchronized. Therefore, $\mathcal{M}_1$ operating in the HRS is capable of either bringing multiple oscillators to a synchronized state or break their synchronization apart. The maximum power consumption of the tuning provided by $I_{\mathcal{M}_1}$ in the HRS is only 0.4~$\mu$W, which can be further suppressed by material engineering for the insulating multilayer. The memristive switching of $\mathcal{M}_2$ is shown in Fig.\ref{fig:4}e. Again, both the VCMA mechanism in the HRS and the effect of the additional current in the LRS, as well as their respective impact on the synchronization, can be clearly observed in Fig.\ref{fig:4}f and g.

\begin{figurehere}
  \begin{center}
  \includegraphics[width=3.6in]{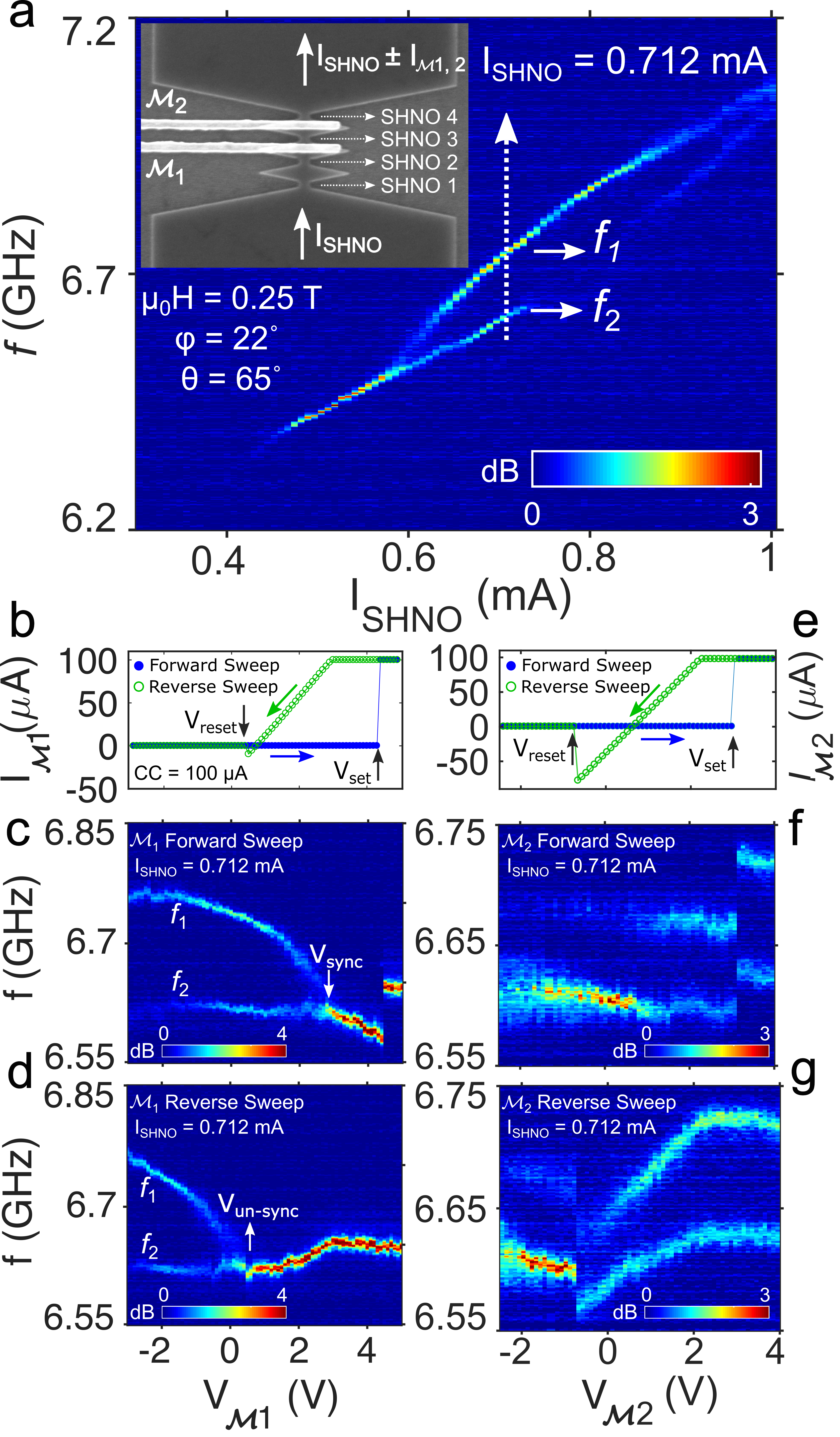}
  \renewcommand{\baselinestretch}{1}
  \caption{\textbf{VCMA dominant control of synchronization in a chain of four SHNOs with two memristive gates.} (a) PSD \emph{vs.}~$I_{\mathrm{SHNO}}$ in a field of $\mu_0 H$~=~0.25 T applied at $\theta$~=~65$^{\circ}$ and $\phi$~=~22$^{\circ}$. The AO starts in a synchronized state, then splits into two branches at higher $I_{\mathrm{SHNO}}$. The inset shows an SEM image of the SHNO with two memristor gates, $\mathcal{M}_1$ and $\mathcal{M}_2$, placed mid-way between the nano-constrictions and the bridges connecting the SHNOs. For $I_{\mathrm{SHNO}}$~=~0.712~mA, (b) $I_{\mathcal{M}_1}$ vs.~$V_{\mathcal{M}_1}$ shows the memristive switching of $\mathcal{M}_1$. (c) PSD \emph{vs.} $V_{\mathcal{M}_1}$ in the forward sweep where the two frequency branches synchronize by VCMA in the HRS. The chain remains synchronized even when $\mathcal{M}_1$ switches to the LRS. (d) PSD response for reverse sweep of $V_{\mathcal{M}_1}$. The chain stays in syncrony until $\mathcal{M}_1$ toggles to the HRS again. (e) $I_{\mathcal{M}_1}$ \emph{vs.} $V_{\mathcal{M}_2}$ and the memristive switching of $\mathcal{M}_2$ having very weak impact on recovering the broken synchronized state neither in (f) forward nor in (g) reverse sweep of $V_{\mathcal{M}_2}$. }
    \label{fig:4}
    \end{center}
\end{figurehere}

Finally, we show how the memristor current itself can dominate over the VCMA effect and hence promote synchronization, or break it reversibly, when the VCMA fails to provide enough coupling to achieve synchronization. Since we have found that we can provide stronger memristor currents in the forward current direction than in the negative, we now operate the same chain as in Fig.~4 but with the direction of the applied field and current reversed. As a consequence, the growing frequency separation vs.~increasing memristor current in Fig.~4g is now replaced by a \emph{decreasing} frequency seperation in Fig.~5e, which will be used to control the mutual synchronization. The positive direction of $I_{\mathrm{SHNO}}$ is  shown by white solid arrows in the inset of Fig.~\ref{fig:5}a. Similar to Fig.~\ref{fig:4}a, the chain starts in synchrony but fails to hold its state and splits into two frequency branches, $f_{3,4}$ and $f_{1,2}$, at more negative $I_{\mathrm{SHNO}}$. Having fixed $I_{\mathrm{SHNO}}$ at --0.757~mA, we swept $V_{\mathcal{M}_1}$ in both the forward and reverse directions and obtained a robust memristive switching as shown in Fig.~\ref{fig:5}b. Note that due to the negative sign of $I_{\mathrm{SHNO}}$, the memristor $I$--$V$ is now pinched at a negative $V_{\mathcal{M}_1}$ value. Fig.~\ref{fig:5}c shows that during the forward sweep, while $V_{\mathcal{M}_1}$ bends the upper frequency branch 
down, it fails to place it within the synchronization bandwidth with the lower 
frequency branch. However, when $\mathcal{M}_1$ toggles to its LRS state at $V_{\mathcal{M}_1}$~=~$V_{\mathrm{set}}$~=~$V_{\mathrm{synch}}$, a noticeable $I_{\mathcal{M}_1}$ flows to the common ground contact through SHNO 3~\&~4, this time opposing $I_{\mathrm{SHNO}}$.  Therefore, the operating current for SHNO 3~\&~4 sharply drops, which is reflected as a sharp decline in the \emph{upper} frequency branch. Since $I_{\mathcal{M}_1}$ only affects SHNO 3~\&~4, we can associate the operating frequency of SHNO 3~\&~4 with the jump in the upper frequency branch, now indicated as $f_{3,4}$. The lower frequency branch then represents SHNO 1~\&~2 and is indicated as $f_{1,2}$. As a result of the sharp frequency jump, the $f_{3,4}$ branch  jumps down to a frequency value near $f_{1,2}$ where they 

\begin{figurehere}
  \begin{center}
  \includegraphics[width=5.5in]{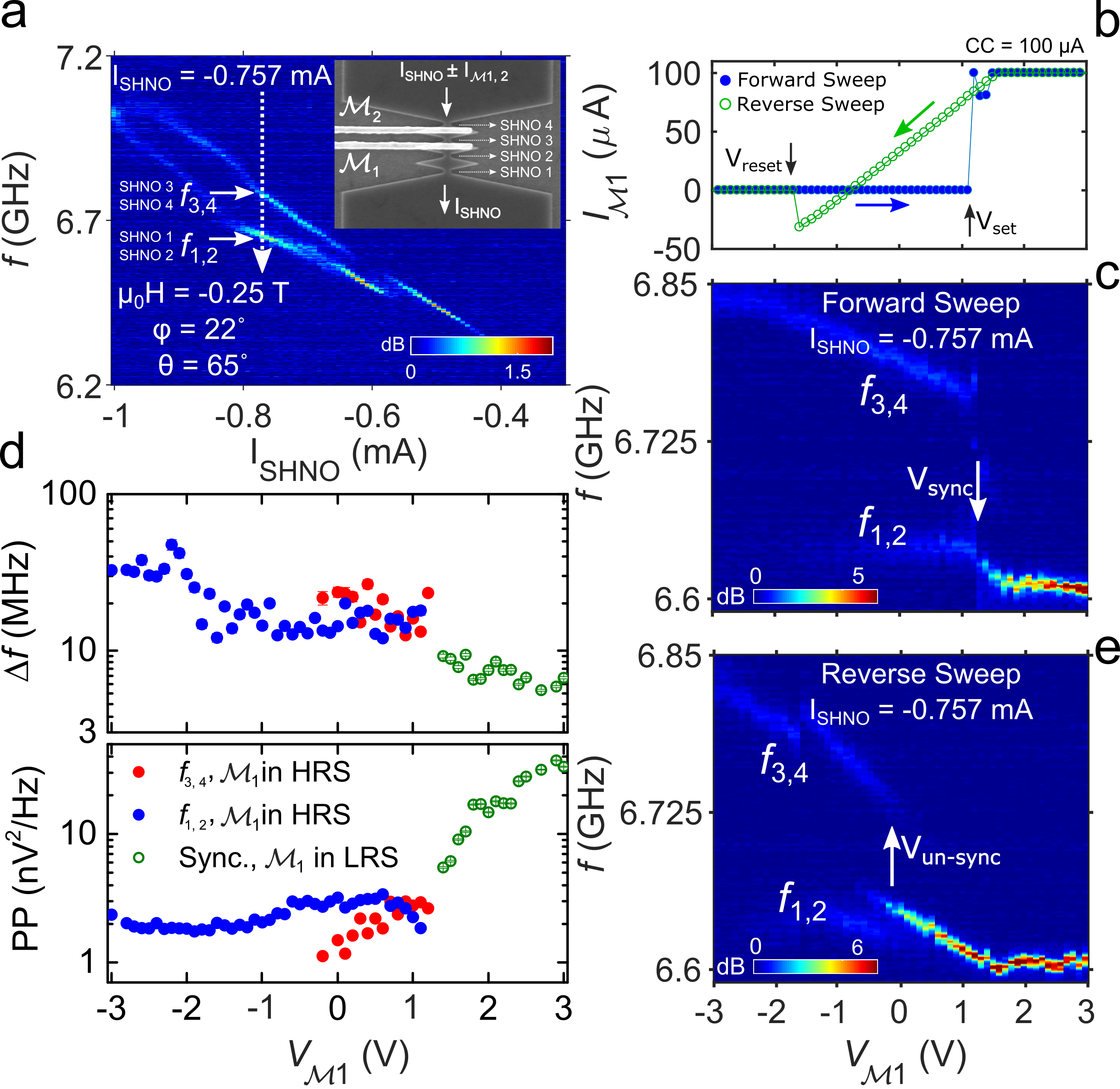}
  \renewcommand{\baselinestretch}{1}
  \caption{\textbf{Memristive dominant control of synchronization in a chain of four SHNOs with two memristive gates.} (a) PSD \emph{vs.}~$I_{\mathrm{SHNO}}$ in a reversed field of $\mu_0 H$~=~-0.25 T applied at $\theta$~=~65$^{\circ}$ and $\phi$~=~22$^{\circ}$. 
  The AO starts in a weak synchronized state, then again splits into two branches at more negative $I_{\mathrm{SHNO}}$. The arrows in the inset define the positive direction of $I_{\mathrm{SHNO}}$. (b)  $I_{\mathcal{M}_1}$ \emph{vs.}~$V_{\mathcal{M}_1}$ showing the memristive switching of $\mathcal{M}_1$. (c) 
  PSD \emph{vs.}~increasing $V_{\mathcal{M}_1}$ at $I_{\mathrm{SHNO}}$~=~--0.757~mA. 
  While the VCMA in the HRS is insufficient to synchronize the SHNOs,  the substantial $I_{\mathcal{M}_1}$ added when $\mathcal{M}_1$ switches to its LRS, sharply pulls the $f_{3,4}$ branch towards $f_{1,2}$ and synchronizes the entire chain. (d) shows how the corresponding extracted linewidth and peak power values improvement when $\mathcal{M}_1$ switches. 
  (e) PSD \emph{vs.}~decreasing $V_{\mathcal{M}_1}$. The synchronized state is first tuned by the decreasing $\lvert I_{\mathcal{M}_1}\rvert$ and then breaks apart when $V_{\mathcal{M}_1}$ switches at 
  $V_{\mathrm{reset}}$.}
  
    \label{fig:5}
    \end{center}
\end{figurehere}

 \noindent synchronize as shown in Fig.~\ref{fig:5}d indicating significant improvement in the peak power and linewith values (hollow green data points). Reversibly, in Fig.~\ref{fig:5}e, when $\mid I_{\mathcal{M}_1}\mid$ is too small at $V_{\mathcal{M}_1}$~=~$V_{\mathrm{un-sync}}$ the synchronized state breaks although $\mathcal{M}_1$ is still in the LRS. The two frequency branches keep increasing their distance when $V_{\mathrm{reset}}$ is met and electric field takes the control back. Again, memristive switching of $\mathcal{M}_2$ shows to have no significant impact on bringing the chain to a synchronized state (see Supplementary Figure 3).

\subsection{Prospect for non-conventional computing.}

Apart from the 
potential of using ultra lower power gating to control and improve the coherency and output power of SHNO arrays as microwave sources, they also hold great potential as emerging computing systems. All coupled oscillatory network-based computing schemes, irrespective of their topology and hardware implementation require a certain input interface, \emph{e.g.}~to provide a bus to feed the data in pattern recognition\cite{Velichko2018electronics, Velichko2019electronics}, map the problem to hardware\cite{parihar2017scirep,chou2019scirep}, and adjust the weights for training purposes in reservoir computing. The latter happens to be a missing piece in recent reservoir computing proposals\cite{zahedinejad2020nt,romera2018nature} for large nanoscale oscillator arrays. 

Memristive control of SHNOs 
provides a scalable approach to both direct input and non-volatile reversible fine-tuning of the individual oscillators and their connections at any location within the network. The memristive approach not only offers such tunability at extremely low power but also opens up a wide range of opportunities to use application-specific memristors from the entire family of memristors\cite{mohammad2016nanotech, li2018jp} depending on the computing paradigm for which the oscillator array is designed. More importantly, the memristive gating approach inherently addresses the memory bottleneck from which conventional von--Neumann processors suffer. In memristor controlled arrays, each processing unit, \emph{i.e.}~each SHNO, has its own memory element and there is hence no need to store and retrieve their states in any external memory limited by data transfer bandwidth. This in turn makes such oscillatory networks with intrinsic memory an appealing candidate for edge computing paradigms that aim to bring computation and data storage closer to the location where it is needed.

\begin{methods}
\subsection{Fabrication of memristive controlled SHNO.}
\sloppy A trilayer stack of W(5)/(Co$_{0.75}$Fe$_{0.25})_{75}$B$_{25}$(1.7)/MgO(2) (thicknesses in nm) was grown at room temperature on an intrinsic high resistivity Si substrate ($\rho_{Si}>$10 k$\Omega \cdot$cm) using an ultra-high vacuum 
sputtering system. DC and RF sputtering were sequentially employed for the depositions of metallic and insulting layers, respectively. The stack was covered with 2~nm thin layer of sputtered AlO$_x$ at room temperature to protect the MgO layer from degradation due to exposure to the ambient conditions. The stack was subsequently annealed at 300~$^{\circ}$C for 1 hour to induce PMA. The stack was then patterned into an array of 4$\times$14$~\mu m^2$ rectangular mesas and the nano-constriction SHNO devices with different widths were defined at the center of these mesas by a combination of electron beam lithography and Argon ion beam etching (IBE) using negative electron beam resist as the etching mask. After removing the electron resist, the sample was covered with 15~nm stoichiometric room--temperature sputtered SiN$_x$ to isolate the gate contacts from the nano--constrcition metallic sidewalls. The gates were defined by sputtering a bilayer of Ti(2~nm)/Cu(40~nm) followed by EBL lithography using negative electron resist to fabricate 100~nm wide gate pattern on top of nano--constriction. The pattern was then transferred to the Ti/Cu bilayer using IBE technique. After removing the remaining negative resist, the sample went through optical lithography using positive resist to define vias in SiN$_x$, giving  access to the SHNO metal layers for the contact pads. Finally, the electrical contacts, including two SG-CPWs for microwave and dc measurements were defined by optical lithography and lift-off for a bilayer of Cu(700~nm)/Pt(20~nm).

\subsection{Memristor characterization and Microwave measurements.}
All microwave electrical measurements were carried out at room temperature using a custom built probe station with the sample mounted at a fixed in-plane angle on an out-of-plane rotatable sample holder between the pole pieces of an electromagnet capable of producing a uniform magnetic field. Using a 6220 Keithley current source, a direct positive electric current, $I_\text{SHNO}$, was made to inject through \textit{dc} port of a high frequency bias-T and the resulting auto-oscillating signal was then amplified by a low-noise amplifier with a gain of +46 dB and subsequently recorded using a spectrum analyzer from Rhode \& Schwarz (10 Hz-40 GHz) comprising a resolution bandwidth of 1 MHz. Two 2400 Keithley source meter were used to sweep the voltage in order to perform the voltage sweeps on the memristors. The output current of the source meter was complied by setting different values for the current compliance on the instrument. Also, the source meter measured the current of the memristive gates for any applied voltage confirm the proper switching between HRS and LRSs (see supplementary Figure 4 for schematic of the measurement setup).




\end{methods}





\begin{addendum}
 \item This work was supported by the Swedish Research Council (VR), the Knut and Alice Wallenberg Foundation, and the Horizon 2020 research and innovation programme (ERC Advanced Grant No.~835068 "TOPSPIN"). The work at Tohoku University was supported by JSPS Kakenhi 17H06093 and 19H05622 and RIEC Cooperative Research Projects.
  \item[Author contributions] S.F. and S.K. and H.O. developed the material stacks. M.Z. designed and fabricated the devices, and carried out all measurements and data analysis. All authors contributed to the interpretation of the results and co-wrote the manuscript.
  \item[Competing Interests] The authors declare that they have no
competing financial interests.
 \item[Correspondence] Correspondence and requests for materials
should be addressed to J. \AA kerman~(email: johan.akerman@physics.gu.se).
\end{addendum}

\clearpage

\beginsupplement
\subsection{\textbf{Supplementary information.} } \label{sup1}

\begin{figurehere}
  \begin{center}
  \includegraphics[width=3.5in]{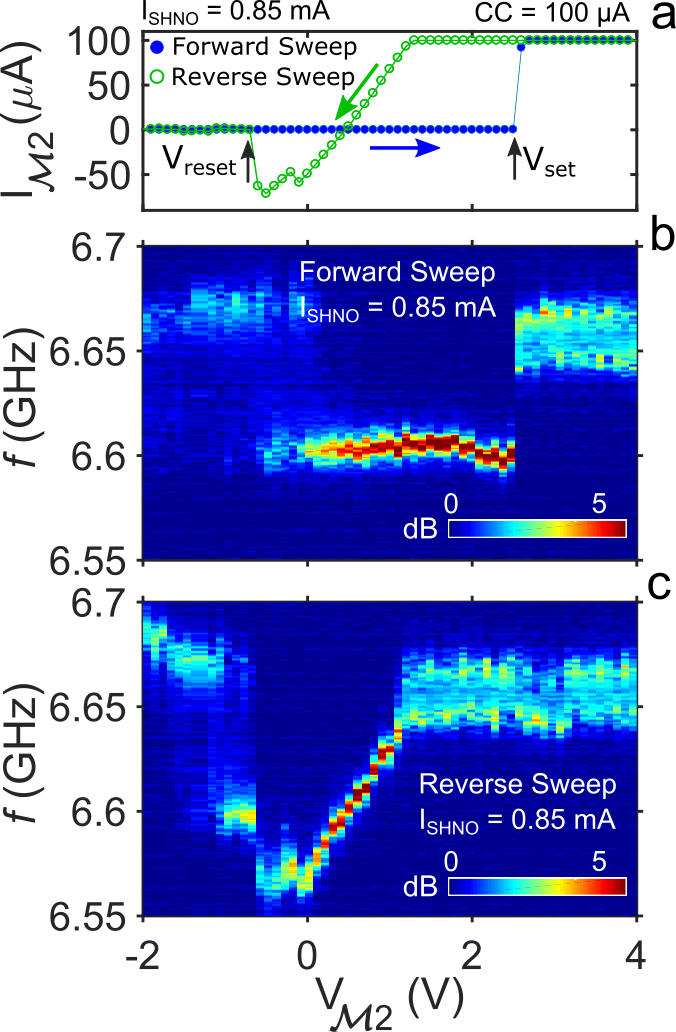}
    \renewcommand{\baselinestretch}{1}
    \caption{\textbf{Supplementary Fig.1. A chain of two SHNOs with two memristive gates. (a)} $I_{\mathcal{M}_2}$ \emph{vs.} $V_{\mathcal{M}_2}$ showing reversible switching between HRS and LRS defined by CC~=~100 $\mu$A for a constant $I_{\mathrm{SHNO}}$~=~0.85~mA. PSD \emph{vs.} $V_{\mathcal{M}_2}$ for \textbf{(b)} the forward and and \textbf{(c)} the reverse sweeps. Similar to sweeping $V_{\mathcal{M}_1}$ in Fig.\ref{fig:2}, ${\mathcal{M}_2}$ in forward sweep provides a strong VCMA tunability to break and restore the synchronized state.
   } 
    \label{fig:supp1}
    \end{center}
\end{figurehere}

\begin{figure}
  \begin{center}
  \includegraphics[width=6in]{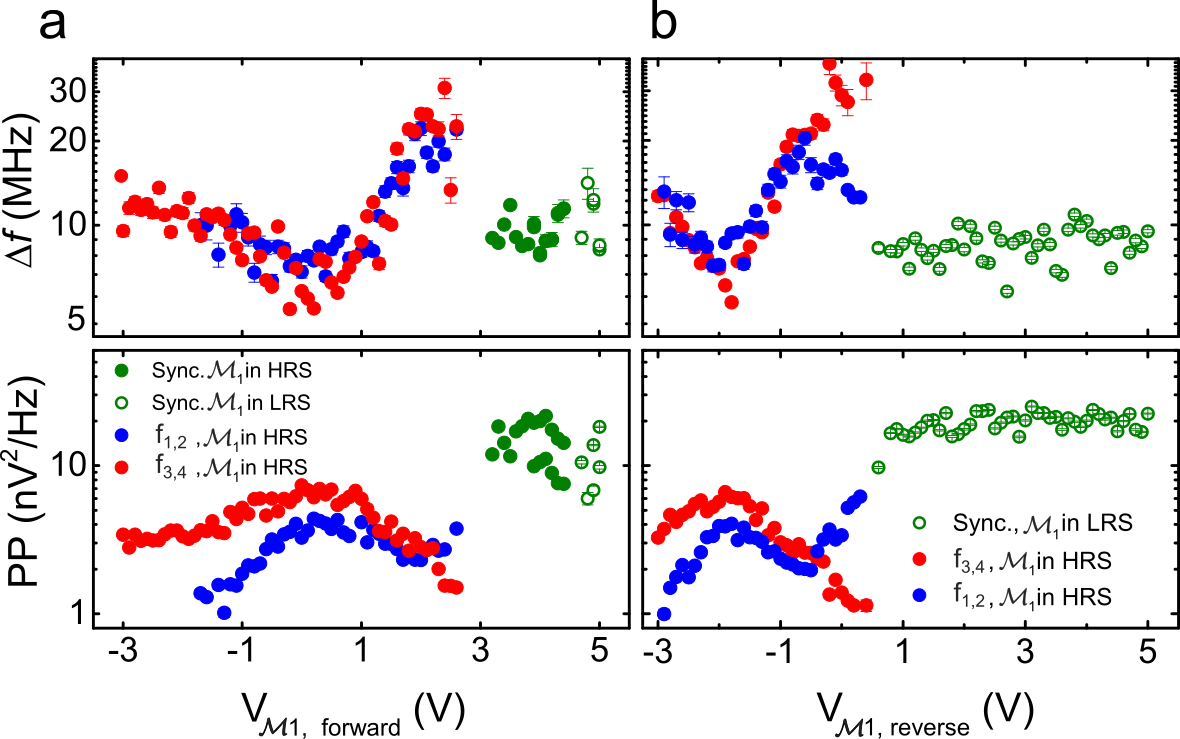}
    \renewcommand{\baselinestretch}{1}
    \caption*{{\textbf{Supplementary Fig.2. Impact of 
    VCMA dominant control on linewidth and peak power values in a chain of four SHNOs. (a)}} Extracted linewidth and peak power values \emph{vs.} $V_{\mathcal{M}_1}$ for the forward sweep in Fig.\ref{fig:4}c. Both linewith and peak power substantially improve when $f_{1,2}$ and $f_{3,4}$ (solid blue and red data points) synchronize at $V_{\mathcal{M}_1}$= $V_{sync}$  while ${\mathcal{M}_1}$ is in the HRS (solid green data points) . The chain remains synchronized even after ${\mathcal{M}_1}$ switches to the LRS (hallow green data points). \textbf{(b)} Extracted linewidth and peak power values \emph{vs.} $V_{\mathcal{M}_1}$ for the reverse sweep in Fig.\ref{fig:4}d. $I_{\mathcal{M}_1}$ keep the SHNOs in synchrony until ${\mathcal{M}_1}$ switches back to the HRS at $V_{un-sync}$ where the VCMA is too weak to keep the chain synchronized.}
    \label{fig:supp2}
    \end{center}
\end{figure}

\begin{figure}
  \begin{center}
  \includegraphics[width=3.5in]{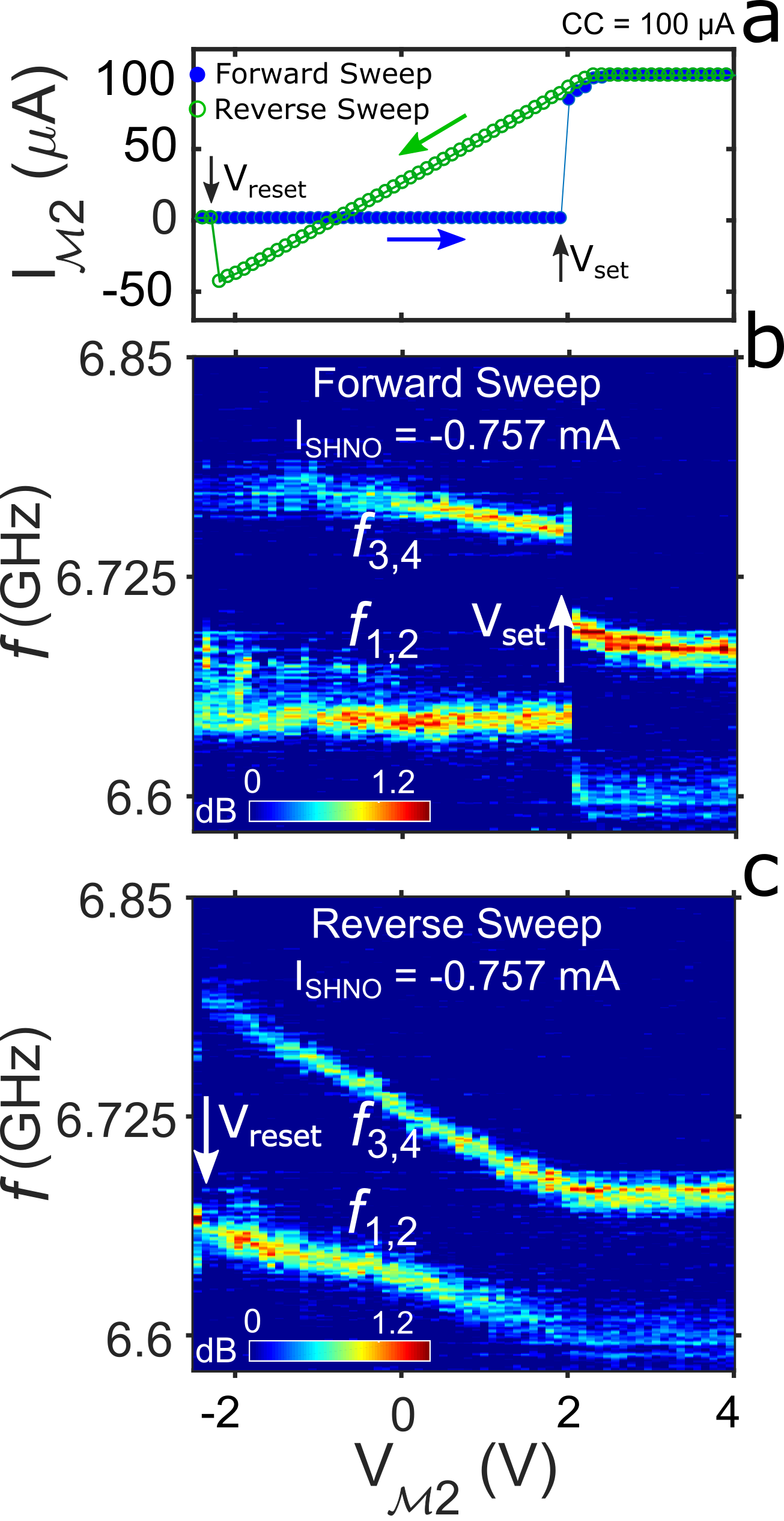}
    \renewcommand{\baselinestretch}{1}
    \caption*{{\textbf{Supplementary Fig.3. Memristive dominant control of synchronization in a chain of four SHNOs with two memristive gates. (a)} } $I_{\mathcal{M}_2}$ \emph{vs.}~$V_{\mathcal{M}_2}$ showing the memristive switching of $\mathcal{M}_2$. (\textbf{b}) PSD \emph{vs.}~increasing $V_{\mathcal{M}_2}$ at $I_{\mathrm{SHNO}}$~=~--0.757~mA. Neither VCMA in the HRS nor the substantial  $I_{\mathcal{M}_1}$ when switching to the LRS are sufficient to synchronize the chain. \textbf{(c)} PSD \emph{vs.}~decreasing $V_{\mathcal{M}_2}$. The chain remains non-synchronized for the entire $V_{\mathcal{M}_2}$ reverse sweep.}
    \label{fig:supp3}
    \end{center}
\end{figure}

\begin{figure}
  \begin{center}
  \includegraphics[width=6in]{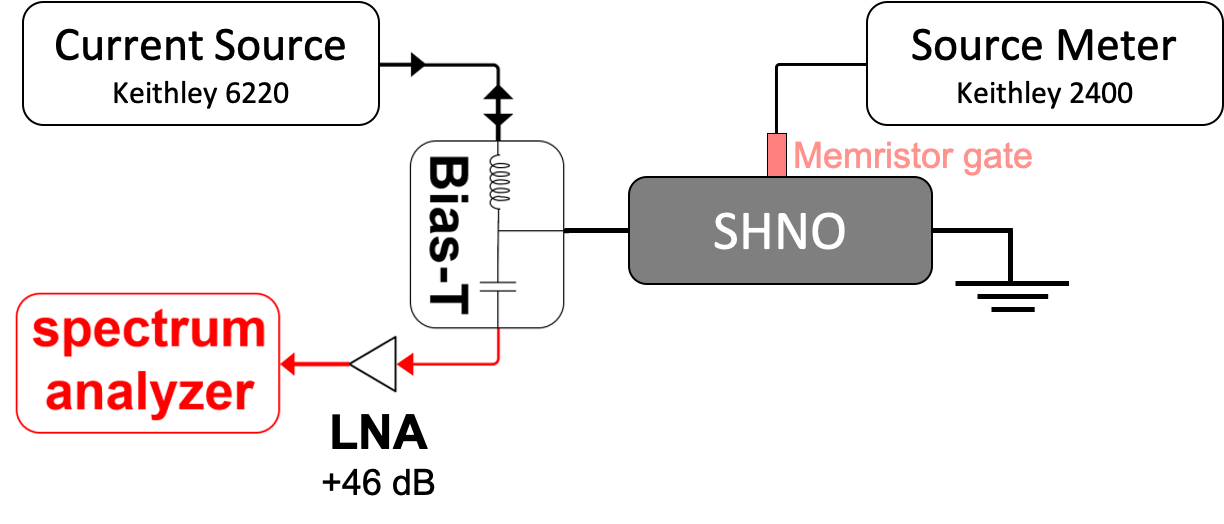}
    \renewcommand{\baselinestretch}{1}
    \caption*{\textbf{Supplementary Fig.4. Schematic of electrical setup for characterizing memristors and microwave measurement of SHNOs.
    } }
    \label{fig:supp3}
    \end{center}
\end{figure}
\clearpage

\end{document}